# Navigating Knowledge Management Implementation Success in Government Organizations: A type-2 fuzzy approach

**Saman Foroutani**
Department of Management and Information Systems, Safashahr Branch, Islamic Azad University, Safashahr, Iran. Email: saman1361@gmail.com
**Nasim Fahimian**
Head of Modular Design, Hamburg, Germany. Email: n.fahimian@gmail.com
**Reyhaneh Jalalinejad**
Department of Economics and Management, University of Padova, Padova, Italy.
Email: reyhane.jalalinejad@gmail.com
**Morteza Hezarkhani**
Department of Economics and Management, University of Padova, Padova, Italy.
Email: moripm.hezar@gmail.com
**Samaneh Mahmoudi**
Department of management
University of Basir, Qazvin, Iran
Email: samanehmahmoudi1991@gmail.com
**Behrooz Gharleghi**
School of Digital Finance, Arden University Berlin, Germany.
Email: gharleghi.bn@gmail.com

## Abstract

Optimal information and knowledge management is crucial for organizations to achieve their objectives efficiently. As a rare and valuable resource, effective knowledge management provides a strategic advantage and has become a key determinant of organizational success. The study aims to identify critical success and failure factors for implementing knowledge management systems in government organizations. This research employs a descriptive survey methodology, collecting data through random interviews and questionnaires. The study highlights the critical success factors for knowledge management systems in government organizations, including cooperation, an open atmosphere, staff training, creativity and innovation, removal of organizational constraints, reward policies, role modeling, and focus. Conversely, failure to consider formality, staff participation, collaboration technologies, network and hardware infrastructure, complexity, IT staff, and trust can pose significant obstacles to successful implementation.

**Keywords:** Knowledge Management, Risk Assessment, Fuzzy Map, Type 2 Fuzzy Numbers, Government Organizations.

## 1. Introduction:

Knowledge is widely recognized as a crucial economic resource and may even be the sole source of competitive advantage for organizations (Drucker, 2009; Levinson & Drucker, 1996). Due to the difficulty in creating, transferring, and replicating knowledge, it holds a strategic position in comparison to other competitive resources within organizations (Ahn & Chang, 2004; Dehkordy et la., 2013; Shamsaddini et al., 2015). Nonaka and Takuchi (1995) distinguish between two types of knowledge in organizations: tacit knowledge, which exists in the form of attitudes, ideas, experiences, and organizational culture, and explicit knowledge, which is found in instructions, regulations, and scientific formulas. They argue that both types of knowledge are transferable (Crossan, 1996; Nonaka & Takeuchi, 1995). Davenport and Prusak (1998) argue that the use of knowledge in organizations can be measured and evaluated, and that organizations require an ongoing tool for assessing the effectiveness and efficiency of their knowledge processes in order to improve their performance. As employees' knowledge or intellectual capital is a vital asset that distinguishes organizations from their competitors, managers must understand how to effectively manage the organization's knowledge (Chung et al., 2016; Forcadell & Guadamillas, 2002; Liebowitz & Suen, 2000; Supyuenyong & Swierczek, 2013; Chen et al., 2021; Hashemzadeh et al., 2011; Gheitarani et al., 2022), and use their intellectual capacities to improve organizational performance. Effective implementation of knowledge management requires attention to key factors that impact the success of system implementation, such as infrastructure (Bakar et al., 2015; Kasemsap, 2015; Gong et al., 2021; Amaravadi, 2005; Pakuhinezhad et al., 2023b). As organizations increasingly recognize the importance of knowledge management, many seek to establish a knowledge management system to facilitate and utilize knowledge management activities. To implement and develop such systems successfully, it is crucial to address the issues, challenges, and factors that impact their success in today's business environment (Ngai & Chan, 2005; Maditinos et al., 2011). Iranian government organizations face unique knowledge management challenges due to cultural and organizational barriers, limited technological infrastructure, knowledge silos and lack of integration, and knowledge retention and succession planning issues. These challenges hinder effective implementation of knowledge management systems. This paper aims to address the following research questions: 1. What are the specific knowledge management challenges faced by government organizations? 2. How do these challenges impact the successful implementation of knowledge management systems within these organizations? 3. What are the key factors that influence the effectiveness of knowledge management systems in the context of Iranian government organizations? 4. How can the type-2 fuzzy map methodology be utilized to identify and analyze these factors? 5. What insights and recommendations can be derived from the empirical investigation to enhance the implementation and utilization of knowledge management systems in Iranian government organizations? Accordingly, this study aims to utilize the type-2 of fuzzy map to identify the factors that affect the successful implementation of knowledge management systems in government organizations in southeast part cities of Iran.

## 2. Literature Review

### 2.1. Knowledge management

In the past decade, attitudes towards organizational resources have undergone a significant shift, possibly due to advances in information technology. Knowledge is now considered a primary resource for creating value and gaining a competitive advantage in the business environment. This shift has led to a greater emphasis on explicit knowledge and a reduced focus on tacit knowledge held by employees and their ability to create knowledge (Liebowitz & Suen, 2000; Nazari et al., 2016). Knowledge management is a process by which an organization can generate wealth and create value from intellectual assets based on its knowledge and intellectual capital (Nonaka & Takeuchi, 1995), due to the rapid and dynamic changes in the organizational environment, managers have realized that information and knowledge is a valuable resource that must be managed because it is very effective in meeting organizational goals (Jalali & Sardari, 2015; Shajera & Al-Bastaki, 2013). Knowledge management has a significant impact on increasing the efficiency and effectiveness of business activities and in other words helps organizations to transfer knowledge and experience to gain a competitive advantage. (Chinying Lang, 2001; Shajera & Al-Bastaki, 2013; Supyuenyong & Swierczek, 2013; Pakuhinezhad, 2023b). So far, various definitions of knowledge management have been presented, which can be generally said: knowledge management is defined as acquiring the knowledge of employees and knowledge outside the organization and using it for further growth and development of the organization (Lee & Choi, 2003; Liebowitz & Frank, 2016; Hakkak et al., 2016). The following are some definitions of knowledge management as expressed by experts:

Davenport and Prasack (1998) believe that knowledge management is the exploitation and development of an organization's knowledge capital to advance the goals of the organization. And knowledge that is managed includes both tacit and explicit knowledge (Anand et al., 2002; Takalo et al., 2013). Mike Borg Knowledge knows, and knowledge management helps individuals communicate with each other and share their knowledge (Sabherwal & Becerra - Fernandez, 2003; Nawaser et al., 2015; Yaghoubi et al., 2011). It flows down the lines of the organization and is rarely available at the right time, but in knowledge-based organizations, knowledge flows throughout the organization and is used by everyone in need (Burk, 1999; Goodman & Darr, 1998). By integrating the three knowledges, effective knowledge management can be achieved. These know-hows are: Knowledge about the customer: This type of knowledge leads to identification and information about customers and the organization can effectively target them. Knowledge for the customer: Using this type of knowledge creates a backbone of knowledge for the customer and makes customers get a better experience of the organization's products and services; And customer knowledge: This type of knowledge refers to thoughts, ideas, ideas and information obtained from customers and provides the basis for optimizing products and services and processes and ideas for typification (J. O. Chan, 2016; Liebowitz & Frank, 2016; Pakuhinezhad et al., 2023c; Gheitarani et al., 2023; Nazari et al., 2016). As a result, the implementation of knowledge management in the organization causes the knowledge produced by individuals to remain in the organization forever and with the departure of employees, the knowledge produced does not leave the organization (Burk, 1999; Tom H Davenport, 2016; Nawaser et al., 2014; Maldonado-Guzmán et al., 2016).

### *Factors affecting the success of knowledge management system implementation*

Identifying and prioritizing critical factors that impact the successful implementation of knowledge management systems is crucial for improving processes. When these factors are considered based on their relative importance, they can have a significant impact on the success of knowledge management implementation (Ansari et al., 2013; Flynn & Arce, 1997; Pakuhinezhad et al., 2023a). This study aims to examine the effect of such factors on the successful implementation of knowledge management systems in government organizations located in the southwestern part of Iran. Zolanski (1996) identified four factors that hinder the transfer of knowledge within organizations: knowledge source, knowledge receiver, nature of knowledge, and knowledge dissemination platform (Zolanski, 1996). Building on Zolanski's model, Smith et al., (2008) discussed the factors influencing knowledge transfer between organizations. They categorized these factors into four categories: knowledge transmitters, knowledge receivers, nature of knowledge, and basis of knowledge dissemination (Smith et al., 2008). Furthermore, Targhen and D'Souza (2010) presented a comprehensive classification of risks associated with knowledge management in organizational networks. They identified five general factors contributing to these risks: nature of cooperation, nature of the network, physical distance, type of activity, and range (Targhen & D'Souza, 2010; Pakuhinezhad, 2016; Dehghanan et al., 2021). These studies provide valuable insights into the challenges and factors affecting knowledge transfer and management within and between organizations. By understanding and addressing these factors, organizations can improve their knowledge sharing practices and enhance their overall performance.

The identification of factors through interviews was not adequate for this study. Instead, to provide a comprehensive description of the dimensions of the problem in the Iranian organizational ecosystem, the study conducted in-depth interviews with experts until reaching theoretical saturation. To illustrate, the table below shows an example of the factors derived from the interviews.

Table 1. Factors derived from the interviews

| Row | Factors | Source |
|-----|---------|--------|
| 1 | Some of the factors that hinder knowledge sharing in organizations are: distrust, fear of mockery and ridicule, undervaluing the expertise and insights of employees, lack of motivation, poor knowledge transfer by senior staff, inadequate support from management, insufficient time allocation, imposition of content and rules, misuse of knowledge for power and influence, omission of context and background information, neglect of new and updated knowledge, ineffective training courses, fear of losing one's job, and fear of criticism. | Interviews with Experts, Heads of Audit, Managers and Senior Experts |

## 2.2. Organizational culture

Organizational culture is one of the most fundamental factors for the success of knowledge management systems, which are the common values and norms in the organization and the employees of the organization, which are the interface between them and determine how things are done in the organization. In other words, organizational culture determines the social identity of any organization (Ansari et al., 2013; Robbins, 1991; Robbins & Butler, 1998), according to the support that culture provides for knowledge management and for knowledge and creation. It values, encourages employees to share and apply knowledge (Ansari et al., 2013; Goh, 2002). Research shows that culture is one of the biggest obstacles facing organizations are in creating a knowledge-based organization (Chase, 1997).

## 2.3. Organizational structure

One of the factors that plays an important role in the application of new technologies in knowledge management is organizational structure and can define a structure for the organization by defining different tasks and coordinating these tasks (Gold & Arvind Malhotra, 2001; Gopalakrishnan & Santoro, 2004; Walczak, 2005) is one of the important indicators in the organizational structure of formality and decentralization and complexity that have a great impact on coordination and cooperation within the company and knowledge creation (Ansari et al., 2013; Lee & Choi, 2003; Sohrabi et al., 2011).

## 2.4. Human resources

Individuals in organizations are considered as human tools including: skills, knowledge roles, motivation and strengthening of learning and creativity networks (Ansari et al., 2013; Moffett et al., 2003; Asghari Moghaddam & Pakuhinezhad, 2024; Pakuhinezhad & Atrian, 2024) on the other hand, knowledge creators in the organization. (Chuang et al., 2016; Inkinen & Inkinen, 2016) while human resources play a key role in knowledge management for a variety of reasons, the main focus is on employee recruitment issues. Is their development and maintenance. Effective staff recruitment is critical and should focus on the ability of volunteers to adapt to the organization's culture in a specific way, rather than adapting to job characteristics (Thomas H Davenport et al., 1998; Rubenstein-Montano et al., 2001; Yahya & Goh, 2002).

## 2.5. IT infrastructure

Undoubtedly, one of the drivers of knowledge is information technology, which can play a set of roles to support knowledge management processes (Lambert et al., 1998; Moghaddam Asghari et al., 2024; Luthra et al., 2016). The role of technology infrastructure in knowledge management, reservoir support Knowledge is the increase of access, knowledge exchange and facilities of the knowledge environment that provides individual, group and organizational interactions and as a tool to help knowledge creation processes in scientific environments (Liebowitz & Frank, 2016; Tan, 2016). In the implementation of knowledge management systems should be considered, the simplicity of the technology used, tailored to the needs of users, the relevance of knowledge content, standardization of knowledge structure and ontology (Migdadi, 2009) without information technology, the possibility of storing information No, and since storage is one of the main processes of knowledge management, weakness in this process leads to inefficiency of the knowledge management system (Lambert et al., 1998; Luthra et al., 2016; Chanda et al., 2024).

## 2.6. Senior management support

An essential principle for success in knowledge management programs is to create a continuous strategic commitment to knowledge production and the support of top and senior managers of the organization, and management in the field of knowledge management must reflect the specific characteristics that lead to knowledge management (Shaqrah, 2016; Gharleghi et al., 2018).

Managers in role modeling reflect knowledge management behavior. And must continually learn and seek new knowledge and ideas (Jafari et al., 2007). Senior managers have an effective role on other success factors in implementing the knowledge management system, such as creating a proper culture, designing training programs and encouraging employees to participate in these programs, and so on. Then, after reviewing the literature and research, and based on the opinion of experts; Factors affecting the success of knowledge management system implementation in 5 dimensions and 18 components were determined and specified, which are shown in Table 1. As shown in this table, the dimensions and components, along with the sources cited, are shown.

**Table 2.** Dimensions and components affecting the success of knowledge management system implementation

| | Dimensions | Abbreviations | Components | Reference |
|---|---|---|---|---|
| **Factors affecting the success of knowledge management system implementation** | Organizational Culture | $x_1$ | Trust | (Aktharsha & Sengottuvel, 2016; Al-Dmour et al., 2016; Thomas H Davenport & Prusak, 1998; Lee & Choi, 2003; Marouf & Agarwal, 2016; Wang & Wang, 2016) |
| | | $x_2$ | Cooperation | |
| | | $x_3$ | Open atmosphere | |
| | | $x_4$ | Knowledge sharing | |
| | | $x_5$ | creativity and innovation | (Thomas H Davenport & Prusak, 1998; Edú-Valsania et al., 2016; Forcadell & Guadamillas, 2002; Singh & Chauhan, 2016) |
| | | | | (Edú-Valsania et al., 2016; Holt et al., 2000; Quin et al., 2005; Siemieniuch & Sinclair, 2004; Singh & Chauhan, 2016) |
| | | | | (Ansari et al., 2013; Mohammadian, 2014; Siemieniuch & Sinclair, 2004) |
| | | | | (Thomas H Davenport et al., 1998; Ho, 2009; Jafari et al., 2007; Kahraman & Tunc Bozbura, 2007; Migdadi, 2009; Pukkila, 2009) |
| | Organizational Structure | $x_6$ | Focus | (Ansari et al., 2013; Forcadell & Guadamillas, 2002; Gaffoor, 2008; Jalaldeen et al., 2009; Lee & Choi, 2003; Siemieniuch & Sinclair, 2004; Walczak, 2005) |
| | | $x_7$ | Recognition | |
| | | $x_8$ | Complication | |
| | | $x_9$ | Role modelling | |
| | | | | (Chinying Lang, 2001; Gaffoor, 2008; Jalaldeen et al., 2009; Lee & Choi, 2003) |
| | | | | (Ansari et al., 2013; Chin Wei et al., 2009; Forcadell & Guadamillas, 2002; Gaffoor, 2008; Siemieniuch & Sinclair, 2004)(Abebe & Kabaji, 2016; Akhavan, 2012; Choy Chong, 2006; Jafari et al., 2007) |
| | Human resources | $x_{10}$ | Employee participation | (Crauise O'Brien, 1995; Jalaldeen et al., 2009; Moffett et al., 2003; Ryan & Prybutok, 2001) |
| | | $x_{11}$ | Employee training | (Cohen & Backer, 1999; Moffett et al., 2003; Ryan & Prybutok, 2001),(Akhavan, |

| | | | | 2012; Al-Mabrouk, 2006; Gai & Xu, 2009; Migdadi, 2009) |
|---|---|---|---|---|
| Information technology infrastructure | $x_{12}$ | Network infrastructure and hardware | | (Ansari et al., 2013; Gaffoor, 2008; Mohammadi et al., 2010; Turban et al., 2008) |
| | $x_{13}$ | Access to applications | | (Lee & Choi, 2003; Turban et al., 2008; Yeh et al., 2006) |
| | $x_{14}$ | IT staff | | (Al-Mabrouk, 2006; Thomas H Davenport et al., 1998; Gai & Xu, 2009; Ho, 2009; Pukkila, 2009) |
| | $x_{15}$ | Collaborative technologies | | |
| Leadership and support of senior managers | $x_{16}$ | Reward policies | | (Gaffoor, 2008; Sunassee & Sewry, 2003; Taylor & Wright, 2004; Yeh et al., 2006),(Al-Mabrouk, 2006; Choy Chong, 2006; Thomas H Davenport et al., 1998; Gai & Xu, 2009; Ho, 2009) |
| | $x_{17}$ | Knowledge strategy | | |
| | $x_{18}$ | Removal of organization restrictions | | |

## 3. Research Methodology

The research employs an applied and descriptive research design, utilizing a one-section and field survey method. The method of data collection is through questionnaires, interviews and library studies. The statistical population of the research consists of governmental organizations in southeast part cities of Iran. Questionnaires were distributed among the study population and after collection, the validity and reliability of the research data collection tool were used to assess the standardization of the research data collection tool. Content validity and construct validity were performed using confirmatory factor analysis. Regarding reliability, Cronbach's alpha value was calculated, which indicated the acceptable reliability of the research questionnaire. Considering the purpose of this research, which is to provide a methodology for explaining the fuzzy map and determining the critical factors of success and failure in government organizations towards better management of factors affecting the implementation of knowledge management, the following steps were considered in this research.

***Step 1: Identify the effective factors:*** In this stage, by in-depth study of research literature and interviews with experts, the effective factors on the implementation of knowledge management in southeast part of Iran' government organizations are identified. This study employs a stratified sampling strategy, which divides the sampling area into different sections and selects a random sample from each subsection. This strategy has several benefits, such as:

• It allows for the examination of population groups that might be overlooked in simple random sampling.

• It enhances the accuracy of the calculations and estimates derived from the statistical sample because the sample size of each subgroup is proportional to the size of the corresponding subgroup.

• It enables the use of a specific sampling method in each subdivision of the statistical population. For instance, cluster sampling can be used instead of random sampling in some subdivisions.

***Step 2: Define verbal expressions and fuzzy numbers of type-2 to measure the identified factors:***
The ambiguity and uncertainty that exists in the human evaluation of these indicators has made the use of definite methods inaccurate and unreliable. Fuzzy logic, considering ambiguity and uncertainty, provides a good tool to deal with them in human evaluations. A brief description of the Type-2 fuzzy logic, which is based on the concept of fuzzy sets, is given below.

Type-2 fuzzy logic: is an extension of fuzzy logic that was introduced by Zadeh (1975). Type-2 fuzzy sets have fuzzy membership degrees, which means they can model and reduce uncertainties more effectively than type-1 fuzzy sets. Type-2 fuzzy sets are also called fuzzy-fuzzy sets, as they provide more information than type-1 fuzzy sets. Type-1 fuzzy sets are a first-order approximation of uncertainty, while type-2 fuzzy sets are a second-order approximation of uncertainty. Type-2 fuzzy sets can also handle linguistic and data uncertainties better than type-1 fuzzy sets, as their membership functions are fuzzy (Mendel, 2007). In this research, type-2 fuzzy sets are used to measure and evaluate the risks of implementing knowledge management, because the knowledge obtained from experts through a questionnaire contains words that are uncertain and quantified with definite numbers, and the nature of knowledge management is qualitative and ambiguous. Therefore, using type-1 fuzzy sets or quantitative data will not yield satisfactory results. Type-2 fuzzy sets can capture the uncertainty and ambiguity of the experts' opinions and the qualitative aspects of knowledge management more accurately.

In the following, the method of preliminary calculations on Type-2 fuzzy is given, if $\tilde{\tilde{A}}_1$ and $\tilde{\tilde{A}}_2$ are two trapezoidal fuzzy numbers of the second type.

$$\tilde{\tilde{A}}_1 = \left(\tilde{A}_1^U, \tilde{A}_1^L\right) = \left(\left(a_{11}^U, a_{12}^U, a_{13}^U, a_{14}^U; H_1(A_1^U), H_2(A_1^U)\right), \left(a_{11}^L, a_{12}^L, a_{13}^L, a_{14}^L; H_1(A_1^L), H_2(A_1^L)\right)\right)$$

$$\tilde{\tilde{A}}_2 = \left(\tilde{A}_2^U, \tilde{A}_2^L\right) = \left(\left(a_{21}^U, a_{22}^U, a_{23}^U, a_{24}^U; H_1(A_2^U), H_2(A_2^U)\right), \left(a_{21}^L, a_{22}^L, a_{23}^L, a_{24}^L; H_1(A_2^L), H_2(A_2^L)\right)\right)$$

آنگاه

$$\tilde{\tilde{A}}_1 + \tilde{\tilde{A}}_2 = \left(\tilde{A}_1^U, \tilde{A}_1^L\right) + \left(\tilde{A}_2^U, \tilde{A}_2^L\right)$$
$$= \left(\left(a_{11}^U + a_{21}^U, a_{12}^U + a_{22}^U, a_{13}^U + a_{23}^U, a_{14}^U\right.\right.$$
$$+ a_{24}^U; min\left(H_1(\tilde{A}_1^U), H_1(\tilde{A}_2^U)\right), min\left(H_2(\tilde{A}_1^U), H_2(\tilde{A}_2^U)\right)\right), \left(a_{11}^L + a_{21}^L, a_{12}^L + a_{22}^L, a_{13}^L + a_{23}^L, a_{14}^L\right.$$
$$\left.\left.+ a_{24}^L; min\left(H_1(\tilde{A}_1^L), H_1(\tilde{A}_2^L)\right), min\left(H_2(\tilde{A}_1^L), H_2(\tilde{A}_2^L)\right)\right)\right)$$

$$\tilde{\tilde{A}}_1 - \tilde{\tilde{A}}_2 = \left(\tilde{A}_1^U, \tilde{A}_1^L\right) - \left(\tilde{A}_2^U, \tilde{A}_2^L\right)$$
$$= \left(\left(a_{11}^U - a_{24}^U, a_{12}^U - a_{23}^U, a_{13}^U - a_{22}^U, a_{14}^U\right.\right.$$
$$- a_{21}^U; min\left(H_1(\tilde{A}_1^U), H_1(\tilde{A}_2^U)\right), min\left(H_2(\tilde{A}_1^U), H_2(\tilde{A}_2^U)\right)\right), \left(a_{11}^L - a_{24}^L, a_{12}^L - a_{23}^L, a_{13}^L - a_{22}^L, a_{14}^L\right.$$
$$\left.\left.- a_{21}^L; min\left(H_1(\tilde{A}_1^L), H_1(\tilde{A}_2^L)\right), min\left(H_2(\tilde{A}_1^L), H_2(\tilde{A}_2^L)\right)\right)\right)$$

$$\tilde{\tilde{A}}_1 \times \tilde{\tilde{A}}_2 \cong \left(\tilde{A}_1^U, \tilde{A}_1^L\right) \times \left(\tilde{A}_2^U, \tilde{A}_2^L\right)$$
$$= \left(\left(a_{11}^U \times a_{21}^U, a_{12}^U \times a_{22}^U, a_{13}^U \times a_{23}^U, a_{14}^U\right.\right.$$
$$\times a_{24}^U; min\left(H_1(\tilde{A}_1^U), H_1(\tilde{A}_2^U)\right), min\left(H_2(\tilde{A}_1^U), H_2(\tilde{A}_2^U)\right)\right), \left(a_{11}^L \times a_{21}^L, a_{12}^L \times a_{22}^L, a_{13}^L \times a_{23}^L, a_{14}^L\right.$$
$$\left.\left.\times a_{24}^L; min\left(H_1(\tilde{A}_1^L), H_1(\tilde{A}_2^L)\right), min\left(H_2(\tilde{A}_1^L), H_2(\tilde{A}_2^L)\right)\right)\right)$$

$$\tilde{\tilde{A}}_1 \div \tilde{\tilde{A}}_2 \cong (\tilde{A}_1^U, \tilde{A}_1^L) \div (\tilde{A}_2^U, \tilde{A}_2^L)$$

$$= \left( \begin{array}{c} \left( \dfrac{a_{11}^U}{a_{24}^U}, \dfrac{a_{12}^U}{a_{23}^U}, \dfrac{a_{13}^U}{a_{22}^U}, \dfrac{a_{14}^U}{a_{21}^U}, ; min\left(H_1(\tilde{A}_1^U), H_1(\tilde{A}_2^U)\right), min\left(H_2(\tilde{A}_1^U), H_2(\tilde{A}_2^U)\right) \right), \\ \left( \dfrac{a_{11}^L}{a_{24}^L}, \dfrac{a_{12}^L}{a_{23}^L}, \dfrac{a_{13}^L}{a_{22}^L}, \dfrac{a_{14}^L}{a_{21}^L}; min\left(H_1(\tilde{A}_1^L), H_1(\tilde{A}_2^L)\right), min\left(H_2(\tilde{A}_1^L), H_2(\tilde{A}_2^L)\right) \right) \end{array} \right)$$

In this paper, fuzzy trapezoidal numbers of the type-2, which are shown in Table 2, are used to evaluate the performance and importance of the factors.

**Table 3.** Linguistic variables of trapezoidal fuzzy numbers of the type-2 (Chen & Lee, 2010)

| Linguistic variables | Trapezoidal fuzzy numbers of type-2 |
|---|---|
| Very Low | ((0، 0، 0، 0/1؛ 1، 1)، (0، 0، 0، /05؛ 0/9، 0/9)) |
| Low | ((0، 0/1، 0/1، 0/3؛ 1، 1)، (0/05، 0/1، 0/1، /02؛ 0/9، 0/9)) |
| Medium | ((0/3، 0/5، 0/5، 0/7؛ 1، 1)، (0/4، 0/5، 0/5، /6؛ 0/9، 0/9)) |
| High | ((0/7، 0/9، 0/9، 1؛ 1، 1)، (0/8، 0/9، 0/9، /95؛ 0/9، 0/9)) |
| Very High | ((0/9، 1، 1، 1؛ 1، 1)، (0/95، 1، 1، 1؛ 0/9، 0/9)) |

***Step 3: Measuring the performance and the degree of importance of the factors:*** to measure the performance and the degree of importance of the qualitative components of the factors affecting the implementation of knowledge management, a questionnaire was prepared and distributed among the studied population. This study applied stratified sampling to address the potential biases and limitations of using interviews and questionnaires. The benefits of this method were discussed in the first step.

***Step 4: Integrating evaluator's opinions:*** various methods such as arithmetic mean, median, and mode can be used to combine the evaluations of multiple decision makers. In this research, the average operator was used to collect experts' opinions, as it is commonly used in research studies.

Assume that the evaluation committee consists of m evaluators ($E_t; t = 1, \ldots, m$ ) and the identified factors are represented by $F_j; j = 1, \ldots, n$ . Also assume that the equation (1) is a trapezoidal type-2 fuzzy number that is used to estimate the verbal expression of the agents' performance, which is assigned to the agent $F_j$ by the evaluator $E_t$, and the equation (2) is a trapezoidal type-2 fuzzy number, which is used to estimate the verbal expression of the importance of the factors, and is assigned to the element $F_j$ by the evaluator $E_t$.

The performance average rank of the fuzzy type 2 $\tilde{\tilde{R}}_j$ and the mean of the importance of the fuzzy type 2 $\tilde{\tilde{W}}_j$, are obtained by gathering the experts' opinions based on relationships (3) and (4).

(1)
$$\widetilde{\widetilde{A}}_{jt} = \left(\widetilde{A}_{jt}^U, \widetilde{A}_{jt}^L\right) = \left(\left(\left(a_{j1t}^U, a_{j2t}^U, a_{j3t}^U, a_{j4t}^U; H1\left(\widetilde{A}_j^U\right) H2\left(\widetilde{A}_j^U\right)\right), \left(\left(a_{j1t}^L, a_{j2t}^L, a_{j3t}^L, a_{j4t}^L; H1\left(\widetilde{A}_j^L\right) H2\left(\widetilde{A}_j^L\right)\right)\right)\right)\right)$$

(2)
$$\widetilde{\widetilde{W}}_{jt} = \left(\widetilde{W}_{jt}^U, \widetilde{W}_{jt}^L\right) = \left(\left(\left(w_{j1t}^U, w_{j2t}^U, w_{j3t}^U, w_{j4t}^U; H1\left(\widetilde{W}_j^U\right) H2\left(\widetilde{W}_j^U\right)\right), \left(\left(w_{j1t}^L, w_{j2t}^L, w_{j3t}^L, w_{j4t}^L; H1\left(\widetilde{W}_j^L\right) H2\left(\widetilde{W}_j^L\right)\right)\right)\right)\right)$$

(3)
$$\widetilde{\widetilde{R}}_j = \left(\widetilde{\widetilde{A}}_{j1} \oplus \widetilde{\widetilde{A}}_{j2} \oplus \ldots \oplus \widetilde{\widetilde{A}}_{jm}\right)/m$$

(4)
$$\widetilde{\widetilde{W}}_j = \left(\widetilde{\widetilde{W}}_{j1} \oplus \widetilde{\widetilde{W}}_{j2} \oplus \ldots \oplus \widetilde{\widetilde{W}}_{jm}\right)/m$$

**Step 5: Diffusion and normalization of fuzzy type 2 weights:** by using equation (5), the fuzzy type 2 performance and significance values are obtained (Abdullah & Zulkifli, 2015).

(5)
$$E(j) = DTraT = \frac{1}{2}\left(\frac{\frac{(U_u - L_u) + (\beta_u.m_{1u} - L_u) + (\alpha_u.m_{2u} - L_u)}{4} + L_u}{+\left[\frac{(U_l - L_l) + (\beta_l.m_{1l} - L_l) + (\alpha_l.m_{2l} - L_l)}{4} + L_l\right]}\right)$$
$$j = 1, 2, \ldots, n$$

In relation (5), DTraT stands for trapezoidal dephasing of Type-2 fuzzy sets. Kahraman et al. (2014) have used it in their research to de-fuzzify trapezoidal fuzzy numbers of the second type. The description of the signs used in this regard is as follows:

$\alpha$ and $\beta$ are the maximum degrees of membership of the lower limit of the second type trapezoidal fuzzy number, and $U_u$ is the largest possible value of the upper limit, $L_u$ is the smallest possible value of the upper limit, and $m_{1u}$ and $m_{2u}$ are the second and third parameters of the upper limit of the fuzzy number. Also, $U_l$ is the largest possible value of the lower limit and $L_l$ is the lowest possible value of the lower limit, and $m_{2l}$ and $m_{1l}$ are the second and third parameters of the lower limit, respectively.

**Step 6: Locating and placing the identified factors on the fuzzy map:** the horizontal axis of this map indicates the performance and its vertical axis indicates the importance of the factors in the evaluated organization. By finalizing the data and dividing each axis into three parts, low, medium and high, 9 places (geographic area) will be formed in this map and each of the factors will appear in a unique place according to the scores obtained in the previous stage. This map, like a geographical map, shows the position of factors in the quality space of the organization. The places located on the diameter of the matrix show the balance between importance and performance, and the elements located in them have a relative balance in their status; Because in these cells, the scores of importance and performance are both average. The upper three places of the main diameter, due to the higher importance of the elements, their performance has not been paid much attention, it indicates the deficiency in the performance of the elements; On the opposite side, the lower three places of this diameter will represent the elements that have an increase in performance in proportion to their importance, meaning that they have performed more than what their priority dictates.

**Step 7: Ranking the critical factors of success and failure in the implementation of knowledge management:** The elements placed in the weakness section are actually the obstacles that currently prevent the implementation of knowledge management in the organization, therefore, they are considered as critical factors of failure. be. Similarly, the elements located in the lower part can be considered as critical success factors. Since it is not possible to pay attention and focus on all these factors for the studied organizations, therefore, at this stage, it is necessary to rank these factors and identify the most important and influential ones.

**(a)** *Determining the balanced performance of indicators:*
considering the importance of elements along with their performance can improve the explanatory power of the success or failure index. With this interpretation, in ranking of the critical factors of success, factors that have higher importance and performance are given more points, and for this purpose, the performance of these elements is multiplied by their importance, and the resulting type 2 trapezoidal fuzzy number is ranked. On the other hand, in order to rank the critical factors of failure, it should be noted that there will be more damaging factors that have lower performance despite their high importance. In order to align these two criteria (performance and importance) and with the aim of achieving an index to determine the criticality of these factors, it is necessary to first subtract the performance of the factors from the number one and then multiply it by the importance. In this way, the resulting index will show the degree of harmfulness of the examined elements. Assume that and the average weight (importance) of the type-2 trapezoidal fuzzy and the average performance (status) of the type-2 trapezoidal fuzzy given to the $j^{th}$ index) $j = 1, 2, \ldots, n($ are in the field of critical factors for the success of knowledge management implementation. The trapezoidal fuzzy numbers of the type-2 indicate the ultimate importance of these success factors, are obtained from equation (6).

$$(6) \qquad CSFS = \sum_{j=1}^{n} \left( \tilde{\tilde{W}}_j \otimes \tilde{\tilde{R}}_j \right)$$

On the other hand, if the scores of critical factors belong to failure, the type-2 trapezoidal fuzzy numbers indicate the level of the impact of these factors on the failure of implementation programs, which is obtained from equation (7).

$$(7) \qquad CFFS = \sum_{j=1}^{n} \left( \tilde{\tilde{W}}_j \otimes \left( 1 - \tilde{\tilde{R}}_j \right) \right)$$

**(b)** *Determining the most influential critical factors of success and failure factors affecting implementation:*
In this stage, the scores obtained from the previous stage are fuzzy ranked to determine the priority of their importance for senior management and planners. Various methods have been proposed for ranking fuzzy numbers. Li and Chen (2008) introduced the concept of ranking the values of type-2 trapezoidal fuzzy sets based on the type-2 "TOPSIS" concept, which provides better results than previous methods. This method ranks fuzzy numbers based on their proximity to positive desirability and distance from negative desirability, and assigns an index as the priority or weight to each fuzzy number. The value of ranking type-2 fuzzy numbers is obtained using equation (8) (E. Chan & Lee, 2008).

(8) $$Rank\left(\overset{\approx}{A_i}\right) = M1\left(\tilde{A_i}^U\right) + M1\left(\tilde{A_i}^L\right) + M2\left(\tilde{A_i}^U\right) + M2\left(\tilde{A_i}^L\right) + M3\left(\tilde{A_i}^U\right) + M3\left(\tilde{A_i}^L\right)$$

$$-\frac{1}{4}\left(S1\left(\tilde{A_i}^U\right) + S1\left(\tilde{A_i}^L\right) + S2\left(\tilde{A_i}^U\right) + S2\left(\tilde{A_i}^L\right) + S3\left(\tilde{A_i}^U\right) + S3\left(\tilde{A_i}^L\right) + S4\left(\tilde{A_i}^U\right) + S4\left(\tilde{A_i}^L\right)\right) + H1\left(\tilde{A_i}^U\right) + H1\left(\tilde{A_i}^L\right) + H2\left(\tilde{A_i}^U\right) + H2\left(\tilde{A_i}^L\right)$$

where in:

(9) $\tilde{A}_i = \left(\tilde{A}_i^U, \tilde{A}_i^L\right) =$
$$\left(\left(a_{i1}^U, a_{i2}^U, a_{i3}^U, a_{i4}^U, H_1(\tilde{A}_i^U), H_2(\tilde{A}_i^U)\right), \left(a_{i1}^L, a_{i2}^L, a_{i3}^L, a_{i4}^L, H_1(\tilde{A}_i^L), H_2(\tilde{A}_i^L)\right)\right)$$

$M_P\left(\tilde{A}_i^j\right)$ Indicates the average $a_{ip}^j$ and $a_{i(p+1)}^j$ which is formed out of relationship (10).

(10) $$M_P^{\left(\tilde{A}_i^j\right)} = \frac{\left(a_{ip}^j + a_{i(p+1)}^j\right)}{2} \qquad 1 \le p \le 3$$

and $Sq\left(\tilde{A}_i^j\right)$ is the standard deviation of [1] $a_{ip}^j$ and $a_{i(p+1)}^j$ which is calculated based on equation (11).

(11) $$S_q\left(\tilde{A_i}^j\right) = \sqrt{\frac{1}{2}\sum_{k=q}^{q+1}\left(a_{ip}^j + a_{i(p+1)}^j\right)^2} \qquad ; \quad 1 \le q \le 3$$

and $Sq\left(\tilde{A_i}^j\right)$ is the standard deviation of $a_{i1}^j$, $a_{i2}^j$, $a_{i3}^j$, and $a_{i4}^j$ which is calculated based on equation (12).

(12) $$S_4\left(\tilde{A_i}^j\right) = \sqrt{\frac{1}{4}\sum_{k=1}^{4}\left(a_{ik}^j - \frac{1}{4}\sum_{k=1}^{4}a_{ik}^j\right)^2}$$

$H_P\left(\tilde{A}_i^j\right)$ indicates the membership value $a_{i(p+1)}^j$ in the trapezoidal membership function.

$$A_i^j \quad , \quad 1 \le p \le 3 \quad , \quad j \in \{U, L\}, \quad 1 \le i \le n$$

$A_i^j \quad , \quad 1 \le p \le 3 \quad , \quad j \in \{U, L\}, \quad 1 \le i \le n$

## 4.Findings:

*First, second and third steps:* after a deep study of the research literature and interviews with experts, the effective factors of identification and during a questionnaire were chosen by the experts, these factors are shown in Table 1. Then, based on the verbal expressions listed in Table 2, the function and degree of importance of each factor was determined.

Content validity: This study measured the content validity using the Lawshe content validity ratio. This ratio requires the opinions of expert experts in the field of the desired test content, who are provided with the objectives of the test and the operational definitions related to the content of the questions. They are asked to rate each question based on a three-point Likert scale of "essential component", "useful but not essential component", and "unnecessary component". Then, the Lawshe content validity ratio is calculated using equation 1.

$$CVR = \frac{n_e - N/2}{N/2}$$

The study identified and finalized the effective dimensions and components of the knowledge management system. Based on this, the study prepared a questionnaire with 18 components and calculated the corresponding CVR. The number of experts was 11 and the CVR for all components was higher than 0.59, indicating that the experts considered all components essential for the implementation of the knowledge management system and that the questionnaire had valid content validity. It also measured reliability using Cronbach's alpha coefficient for all dimensions. The Cronbach's alpha coefficient for all dimensions was above 0.7, suggesting that the questionnaire had acceptable reliability. The table 4 presents the findings of this section.

**Table 4.** Cronbach's alpha

| Dimensions | Cronbach's alpha |
|---|---|
| Organizational Culture | 0.860 |
| Organizational Structure | 0.895 |
| Human Resource | 0.919 |
| Information technology infrastructure | 0.741 |
| Leadership and support of senior managers | 0.790 |

The study also measured reliability using Cronbach's alpha coefficient for all dimensions. The Cronbach's alpha coefficient for all dimensions was above 0.7, suggesting that the questionnaire had acceptable reliability. The table presents the findings of this section.

***Fourth step:*** After collecting the research data using type-2 fuzzy numbers listed in Table 2, the verbal phrases were converted to type-2 fuzzy numbers, and then the expert opinions were integrated using equations (3) and (4). The fuzzy performance average $\tilde{\tilde{R}}_j$( and fuzzy weighted average) $\tilde{\tilde{W}}_j$ ( are shown in Table 3.

**Table 5.** Average rank of fuzzy performance $\tilde{\tilde{R}}_j$ and average importance of fuzzy $\tilde{\tilde{W}}_j$

| | Components | $\tilde{\tilde{W}}_j$ | $\tilde{\tilde{R}}_j$ |
|---|---|---|---|
| $x_1$ | Trust | ((0.333,0.5,0.5,0.7;1,1),(0.417,0.5,0.5,0.6;0.9,0.9)) | ((0.3,0.367,0.367,0.467;1,1),(0.333,0.367,0.367,0.417;0.9,0.9)) |
| $x_2$ | Cooperation | ((0.033,0.133,0.133,0.3;1,1),(0.083,0.133,0.133,0.217;0.9,0.9)) | ((0.2,0.333,0.333,0.5;1,1),(0.267,0.333,0.333,0.417;0.9,0.9)) |
| $x_3$ | open Atmosphere | ((0.033,0.167,0.167,0.367;1,1),(0.1,0.167,0.167,0.267;0.9,0.9)) | ((0,0.1,0.1,0.3;1,1),(0.05,0.1,0.1,0.2;0.9,0.9)) |
| $x_4$ | Knowledge Sharing | ((0.433,0.633,0.633,0.8;1,1),(0.533,0.633,0.633,0.717;0.9,0.9)) | ((0.167,0.267,0.267,0.433;1,1),(0.217,0.267,0.267,0.35;0.9,0.9)) |
| $x_5$ | creativity and innovation | ((0.1,0.2,0.2,0.367;1,1),(0.15,0.2,0.2,0.283;0.9,0.9)) | ((0.2,0.367,0.367,0.567;1,1),(0.283,0.367,0.367,0.467;0.9,0.9)) |
| $x_6$ | Focus | ((0.333,0.467,0.467,0.6;1,1),(0.4,0.467,0.467,0.533;0.9,0.9)) | ((0.033,0.167,0.167,0.367;1,1),(0.1,0.167,0.167,0.267;0.9,0.9)) |
| $x_7$ | Recognition | ((0.4,0.533,0.533,0.667;1,1),(0.467,0.533,0.533,0.6;0.9,0.9)) | ((0.367,0.533,0.533,0.667;1,1),(0.45,0.533,0.533,0.6;0.9,0.9)) |
| $x_8$ | Complication | ((0.467,0.6,0.6,0.733;1,1),(0.533,0.6,0.6,0.667;0.9,0.9)) | ((0.233,0.333,0.333,0.467;1,1),(0.283,0.333,0.333,0.4;0.9,0.9)) |
| $x_9$ | Role modelling | ((0.267,0.433,0.433,0.633;1,1),(0.35,0.433,0.433,0.533;0.9,0.9)) | ((0.033,0.133,0.133,0.3;1,1),(0.083,0.133,0.133,0.217;0.9,0.9)) |
| $x_{10}$ | Employee Participation | ((0.467,0.6,0.6,0.733;1,1),(0.533,0.6,0.6,0.667;0.9,0.9)) | ((0.333,0.467,0.467,0.6;1,1),(0.4,0.467,0.467,0.533;0.9,0.9)) |
| $x_{11}$ | Employee Training | ((0.033,0.167,0.167,0.367;1,1),(0.1,0.167,0.167,0.267;0.9,0.9)) | ((0.267,0.433,0.433,0.633;1,1),(0.35,0.433,0.433,0.533;0.9,0.9)) |
| $x_{12}$ | Network infrastructure and hardware | ((0.333,0.5,0.5,0.7;1,1),(0.417,0.5,0.5,0.6;0.9,0.9)) | ((0.267,0.433,0.433,0.6;1,1),(0.35,0.433,0.433,0.517;0.9,0.9)) |
| $x_{13}$ | Access to applications | ((0.467,0.6,0.6,0.733;1,1),(0.533,0.6,0.6,0.667;0.9,0.9)) | ((0.033,0.1,0.1,0.233;1,1),(0.067,0.1,0.1,0.167;0.9,0.9)) |
| $x_{14}$ | IT staff | ((0.467,0.6,0.6,0.733;1,1),(0.533,0.6,0.6,0.667;0.9,0.9)) | ((0.2,0.333,0.333,0.5;1,1),(0.267,0.333,0.333,0.417;0.9,0.9)) |
| $x_{15}$ | Collaborative technologies | ((0.267,0.433,0.433,0.633;1,1),(0.35,0.433,0.433,0.533;0.9,0.9)) | ((0.467,0.6,0.6,0.733;1,1),(0.533,0.6,0.6,0.667;0.9,0.9)) |
| $x_{16}$ | Reward policies | ((0.1,0.3,0.3,0.5;1,1),(0.2,0.3,0.3,0.4;0.9,0.9)) | ((0.2,0.367,0.367,0.567;1,1),(0.283,0.367,0.367,0.467;0.9,0.9)) |
| $x_{17}$ | Knowledge strategy | ((0.267,0.433,0.433,0.633;1,1),(0.35,0.433,0.433,0.533;0.9,0.9)) | ((0.233,0.367,0.367,0.533;1,1),(0.3,0.367,0.367,0.45;0.9,0.9)) |
| $x_{18}$ | Removal of organization restrictions | ((0.1,0.2,0.2,0.367;1,1),(0.15,0.2,0.2,0.283;0.9,0.9)) | ((0.267,0.433,0.433,0.633;1,1),(0.35,0.433,0.433,0.533;0.9,0.9)) |

***Fifth step:*** The mean diffusive values of the type-2 of fuzzy performance rank $\tilde{\tilde{R}}_j$ (and the mean diffusive values of the type-2 of fuzzy importance $\tilde{\tilde{W}}_j$ (were determined using equation (5). The results can be seen in Table 4.

**Table 6.** The mean diffusive values of the type-2 of fuzzy performance rank $\tilde{\tilde{R}}_j$ and the mean diffusive values of the type-2 of fuzzy importance $\tilde{\tilde{W}}_j$

| Factors | $E\left(\tilde{\tilde{W_j}}\right)$ | $E\left(\tilde{\tilde{R_j}}\right)$ | Factors | $E\left(\tilde{\tilde{W_j}}\right)$ | $E\left(\tilde{\tilde{R_j}}\right)$ |
|---|---|---|---|---|---|
| $x_1$ | 0.494 | 0.364 | $x_{10}$ | 0.585 | 0.455 |
| $x_2$ | 0.143 | 0.331 | $x_{11}$ | 0.175 | 0.429 |
| $x_3$ | 0.175 | 0.116 | $x_{12}$ | 0.494 | 0.423 |
| $x_4$ | 0.611 | 0.273 | $x_{13}$ | 0.585 | 0.110 |
| $x_5$ | 0.208 | 0.364 | $x_{14}$ | 0.585 | 0.331 |
| $x_6$ | 0.455 | 0.175 | $x_{15}$ | 0.429 | 0.585 |
| $x_7$ | 0.520 | 0.514 | $x_{16}$ | 0.293 | 0.364 |
| $x_8$ | 0.585 | 0.331 | $x_{17}$ | 0.429 | 0.364 |
| $x_9$ | 0.429 | 0.143 | $x_{18}$ | 0.208 | 0.429 |

**Sixth step:** According to the data in Table 4, a strategic fuzzy map of the qualitative components affecting the implementation of the knowledge management system in the government organizations was obtained as described in Figure 1. Fuzzy map shows the importance and performance of the factors in the organization. The horizontal axis represents the performance and the vertical axis represents the importance of each factor. The data are divided into three levels: low, medium, and high. The map has nine regions based on these levels, and each factor is placed in a region according to its scores from the previous stage. This map is like a geographical map that shows the positions of the factors in the organization's quality space. The regions on the main diagonal indicate a balance between importance and performance, as the factors in these regions have average scores for both criteria. The top three regions of the diagonal indicate a gap in performance, as the factors in these regions have high importance but low or medium performance. The bottom three regions of the diagonal indicate an excess in performance, as the factors in these regions have low or medium importance but high performance. This map shows the importance and performance of the factors in the organization. The horizontal axis represents the performance and the vertical axis represents the importance of each factor. The data are divided into three levels: low, medium, and high. The map has nine regions based on these levels, and each factor is placed in a region according to its scores from the previous stage. This map is like a geographical map that shows the positions of the factors in the organization's quality space. The regions on the main diagonal indicate a balance between importance and performance, as the factors in these regions have average scores for both criteria. The top three regions of the diagonal indicate a gap in performance, as the factors in these regions have high importance but low or medium performance. The bottom three regions of the diagonal indicate an excess in performance, as the factors in these regions have low or medium importance but high performance.

**Seventh step:** Ranking the critical factors of success and failure, effective factors on the implementation of knowledge management system. By using the relationships (6) and (7), the critical success factors and critical failure factors, effective factors on the implementation of knowledge management system are shown in Table 5.

**Table 7.** Determining the critical success and failure factors

| Critical failure factors | | Factors | Critical success factors | | Factors |
|---|---|---|---|---|---|
| $\left(\overset{\approx}{W}_j \otimes \left(1 - \overset{\approx}{R}_j\right)\right)$ | | | $\left(\overset{\approx}{W}_j \otimes \overset{\approx}{R}_j\right)$ | | |
| 0.556,0.733,0.),(1,1;0.429,0.733,0.733,1.4)) ((0.9,0.9;733,1 | | $x_1$ | ((0.007,0.044,0.044,0.15;1,1),(0.022,0.044,0.044,0.09;0.9,0.9)) | | $x_2$ |
| 0.9,0.9;0.75,1,1,1.286),(1,1;0.55,1,1,1.667)) (( | | $x_7$ | ((0,0.017,0.017,0.11;1,1),(0.005,0.017,0.017,0.053;0.9,0.9)) | | $x_3$ |
| 0.425,0.556,0.5),(1,1;0.318,0.556,0.556,1)) ((0.9,0.9;56,0.75 | | $x_8$ | ((0.02,0.073,0.073,0.208;1,1),(0.043,0.073,0.073,0.132;0.9,0.9)) | | $x_5$ |
| 0.6,0.778,0.),(1,1;0.455,0.778,0.778,1.286)) ((0.9,0.9;778,1 | | $x_{10}$ | ((0.011,0.078,0.078,0.22;1,1),(0.04,0.078,0.078,0.142;0.9,0.9)) | | $x_6$ |
| 0.583,0.867,0.),(1,1;0.381,0.867,0.867,1.8)) ((0.9,0.9;867,1.24 | | $x_{12}$ | 0.029,0.058,0.0),(1,1;0.009,0.058,0.058,0.19)) ((0.9,0.9;58,0.116 | | $x_9$ |
| ((0.273,0.556,0.556,1.071;1,1),(0.4,0.556,0.556,0.781;0.9,0.9)) | | $x_{14}$ | 0.035,0.072,0.),(1,1;0.009,0.072,0.072,0.232)) ((0.9,0.9;072,0.142 | | $x_{11}$ |
| ((0.737,1.385,1.385,2.75;1,1),(1,1.385,1.385,1.905;0.9,0.9)) | | $x_{15}$ | 0.057,0.11,0.11,0.),(1,1;0.02,0.11,0.11,0.283)) ((0.9,0.9;187 | | $x_{16}$ |
| | | | 0.053,0.087,0.),(1,1;0.027,0.087,0.087,0.232)) ((0.9,0.9;087,0.151 | | $x_{18}$ |

Considering the points obtained from the previous stage for the critical factors of success and failure, the ranking value of these numbers was determined using equation (8) as described in Table 6.

**Table 8.** Ranking of critical success and failure factors

| Ranking of critical failure factors | | | | Ranking of critical success factors | | | |
|---|---|---|---|---|---|---|---|
| Row | | Indicators | Rank | Row | | Indicators | Rank |
| 1 | $x_7$ | Recognition | 5.423 | 1 | $x_2$ | Cooperation | 51.237 |
| 2 | $x_{10}$ | Employee Participation | 5.411 | 2 | $x_3$ | open Atmosphere | 43.549 |
| 3 | $x_{15}$ | Collaboration technology | 5.297 | 3 | $x_{11}$ | Employee Training | 43.549 |
| 4 | $x_{12}$ | Network infrastructure and hardware | 5.043 | 4 | $x_5$ | creativity and innovation | 33.018 |
| 5 | $x_8$ | Complication | 4.96 | 5 | $x_{18}$ | Removal of organization restrictions | 33.018 |
| 6 | $x_{14}$ | IT staff | 4.952 | 6 | $x_{16}$ | Reward policies | 24.562 |
| 7 | $x_1$ | Trust | 4.869 | 7 | $x_9$ | Role Modelling | 17.383 |
| | | | | 8 | $x_6$ | Focus | 16.472 |

This study proposes a new method for evaluating the qualitative elements that affect the implementation of knowledge management systems in government organizations, using the gap model and a type-2 of fuzzy approach. The method helps identify basic solutions for preventing system failure and deterioration, and promotes a movement towards excellence and improved performance. By appropriately managing ambiguity and uncertainty in the evaluation, this method offers more reliable results compared to other methods.

## 5. Conclusion and Discussion:

In today's organizations, knowledge management systems are increasingly sought after as a means of facilitating knowledge management activities and realizing their benefits. As such, managers place special emphasis on the effective implementation and development of these systems, taking into account the various issues, challenges, and factors that contribute to their success (Ngai & Chan, 2005; Maditinos et al., 2011). Therefore, organizations should integrate knowledge management systems to collect, categorize, organize, store, share, and make knowledge available at the organizational level, thus ensuring that knowledge produced by individuals remains within the organization indefinitely (Burk, 1999; Davenport, 2016; Maldonado-Guzmán et al., 2016). However, human evaluations are prone to inaccuracy and ambiguity, which can undermine social science research. Fuzzy logic offers a suitable tool for handling ambiguity and uncertainty in human evaluations, with various verbal expressions and membership functions proposed for this purpose (Salojärvi et al., 2010). In this research, a type-2 fuzzy approach has been employed to implement knowledge management systems in government organizations, specifically addressing uncertainty and ambiguity in human evaluations.

To achieve our aim, we first identified and evaluated the effective factors on knowledge management system implementation in government organizations using verbal expressions and type-2 fuzzy numbers. This resolved the first issue of gap measurement based on inaccurate evaluations. However, evaluation alone is insufficient without proposing an improved solution. We then placed the identified qualitative elements in a fuzzy qualitative map to identify their strengths and weaknesses. This map provides an overview of the system's situation, highlighting factors such as cooperation, an open atmosphere, employee training, creativity and innovation, removal of organizational limitations, reward policies, modeling, and concentration as critical to success. Conversely, factors such as formality, employee participation, collaboration technologies, network and hardware infrastructures, complexity, information technology employees, and trust are critical to failure. Addressing these factors can help government organizations achieve their goals more efficiently and maximize benefits. Additionally, the vital factors of failure should be considered as decisive points in the improvement strategy of the organization. In the following section, we present solutions and suggestions based on the critical factors of failure.

*Knowledge management and employee participation:* Participation can be defined as a set of workflows and operations that involve all employees in an organization's decision-making process and make them partners. It creates the perception that people participate in knowledge management activities within the organization. Therefore, government organizations should emphasize employee participation in knowledge management, including voluntary cooperation in knowledge storage, applying and sharing knowledge at the organizational level, and contributing ideas, opinions, and initiatives to problem-solving. The basis of this process is the division of authority between management and employees, with management creating a system and atmosphere that encourages employee cooperation and participation in decision-making and problem-solving using organizational knowledge.

*Information technology and knowledge management:* Knowledge can be created through informal and self-organized networks within companies that gradually become systematic, often through face-to-face conversations, phone calls, or email and communication networks. Advances in information technology have created new possibilities for knowledge management, such as decision support systems (DSS) and electronic performance support systems (EPSS). With an increasing number of personal computers and communication networks, organizations can gain better competitive positions by acquiring and maintaining new knowledge. Computer networks enable communication between people with common goals, regardless of geographic location, and facilitate the sharing and combination of ideas and creativity across boundaries of time and space. The most significant value of information technology in knowledge management is its ability to increase access to knowledge and accelerate its transfer. Additionally, government organizations should consider that information technology enables the extraction of knowledge from knowledge holders' minds, which can be organized into regular formats and transferred to internal members and external business partners worldwide.

*Creating a collaborative culture:* Knowledge management encompasses cultural changes in the way developed knowledge is understood, beyond software technology. Successful implementation of knowledge management requires creating a culture of knowledge sharing in the organization, which may involve redesigning organizational values and incentives for participation to achieve organizational goals. Managers should evaluate employee performance based on their efforts, recognizing that cultural changes take time. Organizations should actively participate in knowledge management projects, maintain the knowledge management system, and invest in organizational learning to optimize its effectiveness.

*Reward and motivation:* Incentive programs in knowledge management should encourage employees to achieve goals, improve performance, and increase participation. Programs can include performance evaluations, salary increases, job promotions, financial rewards, recognition, and ranking based on competition.

## Limitations and future research suggestions

This research has some limitations that warrant further investigation. First, it focused on government organizations, which may differ from other organizations in terms of characteristics and challenges. Thus, the findings may have limited generalizability. Future research could apply the type-2 fuzzy approach to other sectors and industries, such as private, non-profit, or educational ones, and compare the results. Second, it used only one type-2 fuzzy method (DTraT) to defuzzify the type-2 fuzzy numbers. Future research could explore other type-2 fuzzy methods, such as the centroid or the Karnik-Mendel methods, and examine their effects on the results. Third, it did not consider the interrelationships and dependencies among the factors affecting knowledge management system implementation. Future research could use techniques such as fuzzy cognitive maps or fuzzy DEMATEL to model and analyze the causal relationships among the factors. Fourth, it did not evaluate the impact of knowledge management system implementation on organizational performance and outcomes. Future research could use measures such as ROI, customer satisfaction, innovation, or productivity to assess the benefits of knowledge management system implementation in government organizations. Fifth, it relied on subjective interpretations of the phenomenon under study, which may vary among different researchers and stakeholders. Future research could use multiple sources of data and methods of analysis to enhance the validity and reliability of the findings.